\documentclass[a4,11pt,twocolumn]{paper}
\usepackage[T1]{fontenc}
\usepackage{graphicx}
\usepackage{listings}
\usepackage{hyperref}
\usepackage{amsmath}
\usepackage{bbding}
\usepackage{orcidlink}
\usepackage{subcaption}
\usepackage{authblk}

\begin{document}
\title{Comparative Analysis of Procedural Planet Generators}

\author[1]{Manuel Zechmann}
\author[1]{Helmut Hlavacs}
\affil[1]{University of Vienna, Faculty of Computer Science

Education, Didactics and Entertainment Computing Research Group\\
  \tt manuel.zechmann@gmail.com, helmut.hlavacs@univie.ac.at
 }

\graphicspath{ {./figures/} }

\date{}

\maketitle

\begin{abstract}
This paper presents the development of two distinct real-time procedural planet generators within the Godot engine: one employing Fractal Brownian Motion (FBM) with Perlin Noise, and another adapting Minecraft-inspired layered noise techniques. We detail their implementation, including a quadtree-based Level of Detail (LOD) system and solutions for planetary mesh generation. A comparative user study (N=15) was conducted where participants explored unique instances generated by our two algorithms alongside two existing procedural planet projects. 
\end{abstract}

\section{Introduction}

Procedural generation enables the creation of vast, diverse worlds without the need for enormous storage. This technique is widely used in video games, such as ``No Man's Sky''~\cite{NoMansSky} and ``Minecraft,''~\cite{Minecraft} where algorithms generate massive environments, like unique planets or expansive terrains, with relatively small data sizes. Beyond gaming, procedural generation also finds applications in fields like simulations, urban planning, and real-time modeling, where efficient creation of large and dynamic environments is essential~\cite{ProceduralPlanning}.

This paper details the design, implementation, and comparative evaluation of two distinct real-time procedural planet generation approaches within the Godot game engine. A key challenge in real-time planetary rendering is balancing visual fidelity with performance, particularly given the complexities of spherical geometry. To address this, our framework incorporates a quadtree-based Level of Detail (LOD) system that dynamically adjusts terrain complexity based on camera proximity, alongside solutions for mesh seam stitching and managing large-scale floating-point precision issues by recompiling Godot for 64-bit precision.

A comparative study (N=15) assessed how these methods, and two reference projects, influenced immersion, exploration, realism, and controls, with participants exploring unique instances from our generators.

\section{Related Work}

Procedural Content Generation (PCG) is a key technique for efficiently creating vast and detailed worlds in applications such as video games and film. With the growing scale and complexity expected in these areas, PCG offers significant advantages by reducing manual asset creation workload, allowing design teams to focus on core gameplay or narrative elements. PCG particularly excels in generating large-scale environments, including peripheral areas that contribute to a sense of scale and realism with minimal authoring effort~\cite{ProceduralGenerationBasics}.

Noise functions are fundamental to procedural content generation, providing the pseudo-randomness necessary for creating diverse and naturalistic patterns in terrains, textures, and other generated content. A wide array of noise functions exists, from simple White Noise to more structured types like lattice gradient noises (e.g., Perlin Noise~\cite{Perlin1985} and Simplex Noise) and explicit noises (e.g., Wavelet Noise), each offering distinct visual characteristics and performance trade-offs. An extensive comparison of common procedural noise functions, detailing their computational efficiency, dimensional scalability, visual quality, and memory requirements, is provided by Lague et al.~\cite{ProceduralNoiseFunctions}. 

The application of Procedural Content Generation (PCG) spans diverse content types in interactive entertainment. De Carli et al.~\cite{ProceduralContentGenerationTechniques} provide a comprehensive survey, categorizing PCG techniques by content (e.g., terrain, cities) and the level of human involvement required, offering a valuable taxonomy for developers. Noise-based PCG has been particularly successful in generating vast game worlds. ``Minecraft'' stands as an iconic example, utilizing multiple layers of Perlin noise to create its expansive block-based terrains and diverse biomes~\cite{MinecraftTerrainGeneration, MinecraftWiki}, inspiring aspects of our "Minecraft-inspired Planet" generator's layered approach. 

Generating realistic and engaging terrain is a cornerstone of creating immersive virtual worlds, especially for expansive planetary environments. A multitude of procedural techniques have been explored, broadly ranging from direct noise application to more complex simulation-based and learning-based methods. Noise-based algorithms, particularly those employing Perlin noise and Fractal Brownian Motion (FBM), are widely used for their ability to produce naturalistic, fractal landscapes. Li and Gao~\cite{FBMTerrain}, for instance, demonstrated a method using FBM with modified random midpoint displacement for controllable and diverse terrain features, an approach sharing foundational principles with our ``Simple Planet'' generator. 

Creating complete procedural planets extends beyond terrain height generation to encompass atmospheric effects, biome distribution on a spherical surface, and overall environmental coherence. While practical guidance for implementing planetary features such as spherical mesh construction, LOD systems, and basic biome layers can be found in community resources and tutorials (e.g., inspired by work from Lague~\cite{YouTubeTutorialSebastianLague} and SimonDev~\cite{YouTubeTutorialSimonDev}), academic research also addresses these complex integrations. 

Efficiently rendering expansive planetary terrains necessitates robust Level of Detail (LOD) systems to balance visual fidelity with real-time performance. Quadtree-based structures are a common approach for dynamically adjusting mesh complexity based on viewer proximity, particularly for large, regular grid terrains, as explored by Liang, Zhou, and Liu~\cite{LargeScale3DTerrain}. They emphasized that for very large scales, basic LOD needs to be augmented with further optimizations like view frustum culling and adaptive tiling to manage memory and rendering load effectively. 

\section{Creation of the Planet Generators}

This work details the development of two procedural planet generation frameworks, which allow for creation of a random new planet on demand. The complete project is publicly available on GitHub \url{https://github.com/mani1709/PlanetGenerator}
and the thesis containing more information is available on the website of the University of Vienna\footnote{\url{https://phaidra.univie.ac.at/detail/o:2124844}}. 
The engine used for this project is Godot, an open-source game engine known for its flexibility, ease of prototyping, and modifiability. To address precision issues rooted in the large scales involved in planetary generation, the engine was adapted to use 64-bit floating-point precision instead of 32-bit by compiling it from source with the \texttt{precision=double} flag~\cite{godotCompile}. This adjustment ensures accurate position calculations at planetary scales while maintaining the benefits of Godot’s rapid development environment.

The first of the two planet generators (called ``Simple Planet'', see Figure~\ref{fig:our_generated_planets} (a)) used in this paper employs Fractal Brownian Motion (FBM) with Perlin Noise to generate 2D heightmaps. By combining multiple octaves of noise, this method produces naturalistic fractal landscapes. Key FBM parameters such as persistence, lacunarity, exponentiation, and the number of octaves are configurable; by adjusting these, developers can generate a wide range of distinct planetary terrain styles, from smooth hills to mountainous worlds. Heights are scaled by a base factor and clamped to always be above a specified ocean level, ensuring flat ocean surfaces. 

The second planetary generator (``Minecraft-inspired Planet'', see Figure~\ref{fig:our_generated_planets} (b)) adapts layered noise for more intricate terrain and biome control, drawing inspiration from games like Minecraft but tailored for spherical surfaces. This method utilizes four distinct 2D Perlin noise layers to generate values for continentalness, erosion, peaks \& valleys, and a random temperature component. A key feature for control is the use of configurable spline curves for each noise generator. These splines allow developers to define custom remapping functions by setting control points (e.g., specifying that raw noise inputs from 0.1-0.3 should be remapped to output values of 0.4-0.5). 

The final terrain height is dynamically calculated based on these noise layers. The continentalness and peaks \& valleys values are combined, and this sum is then multiplied by the inverse erosion value (where higher erosion reduces the height). This resulting height factor is scaled by a configurable maximum terrain amplitude and applied to displace each vertex of the normalized quadsphere along its normal, effectively sculpting the planetary surface. An ocean level parameter also ensures a minimum planetary radius, creating consistent water bodies.

\begin{figure}[hbt!]
  \centering
  \begin{subfigure}[b]{0.48\textwidth}
    \includegraphics[width=\textwidth]{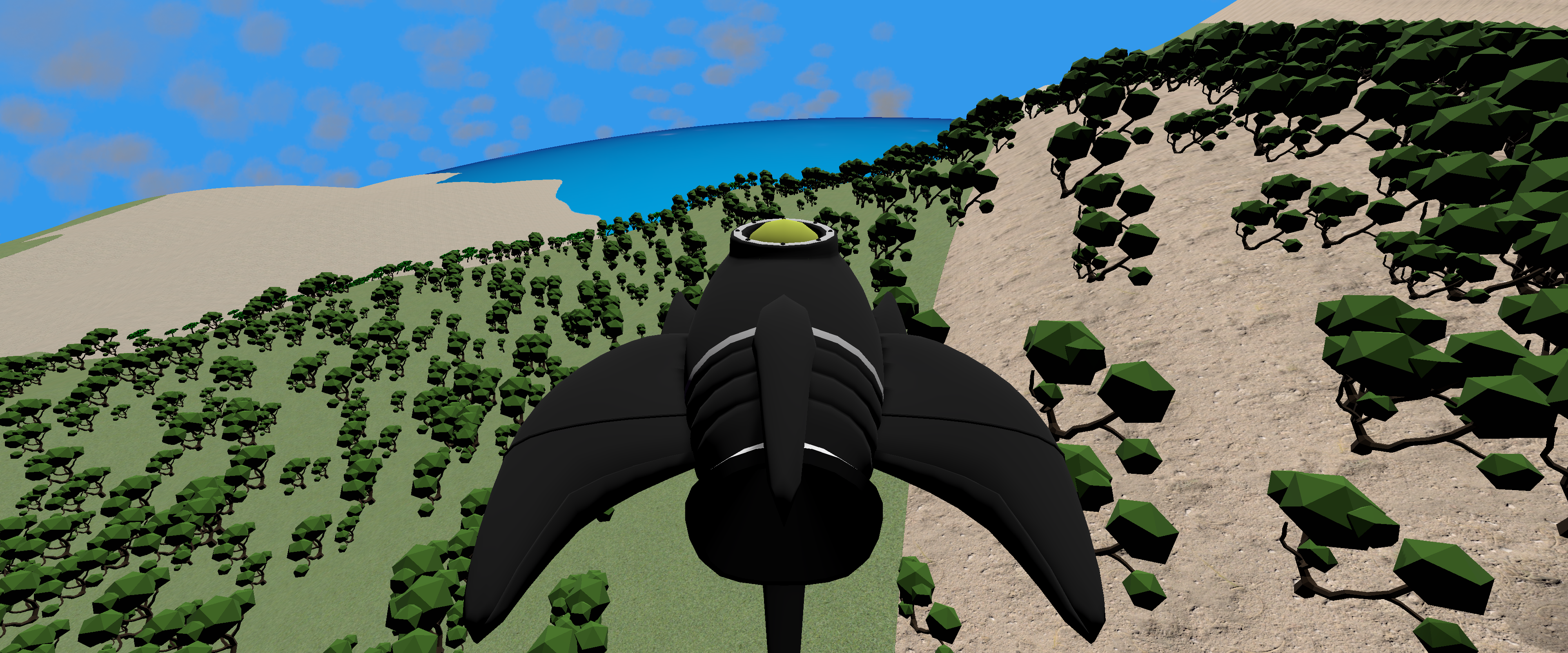} 
    \caption{The ``Simple Planet''.}
    \label{fig:simple_planet_visual}
  \end{subfigure}
  \hfill
  \begin{subfigure}[b]{0.48\textwidth}
    \includegraphics[width=\textwidth]{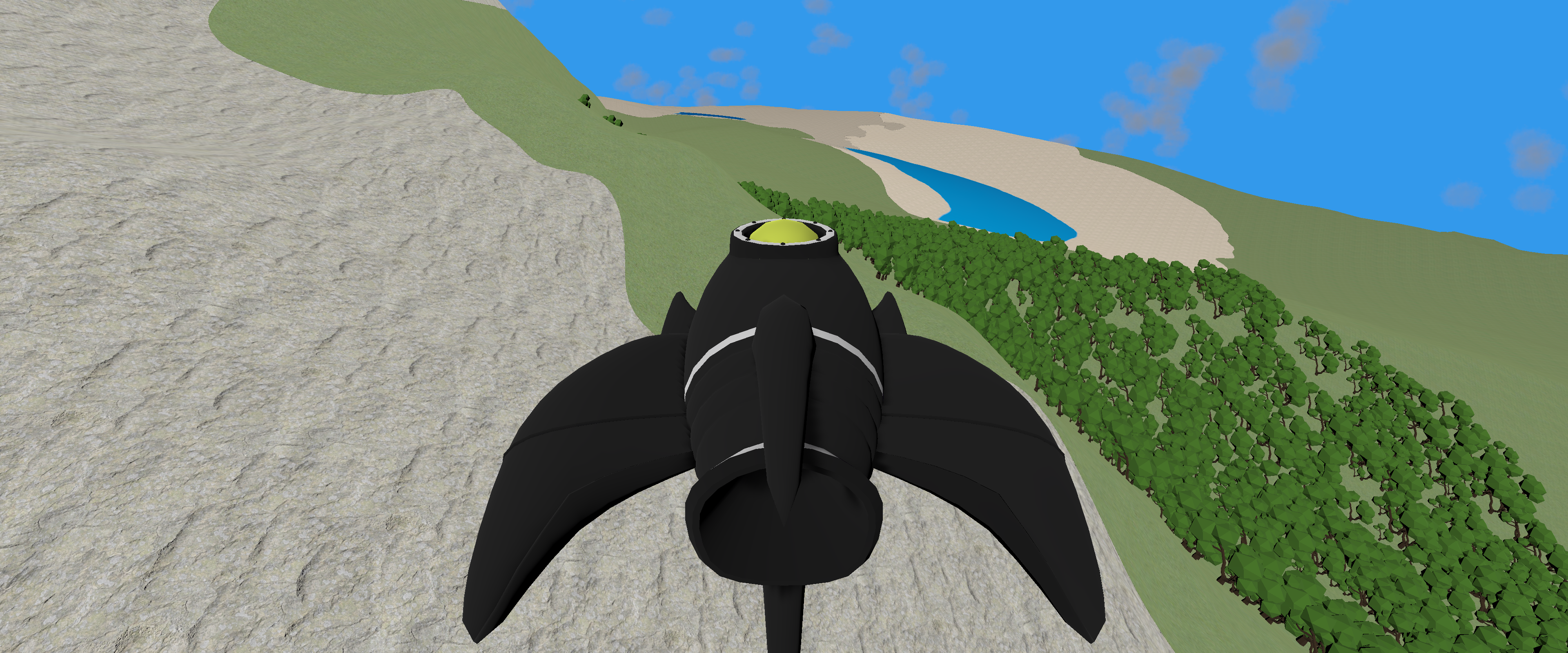}
    \caption{The ``Minecraft-inspired Planet''.}
    \label{fig:minecraft_planet_visual}
  \end{subfigure}
  \caption{Example planets for the two procedural planet generators}
  \label{fig:our_generated_planets}
\end{figure}

Water and lava surfaces are rendered using dynamic shaders at higher LODs, reverting to simple base colors at lower LODs to optimize performance. These shaders simulate depth, flow, and wave motion using blended normal maps and adjustable parameters. Fresnel effects enhance reflections, and the shaders integrate seamlessly with terrain by blending textures near shorelines based on height, creating realistic transitions. The lava shader uses similar logic but with faster animations and volcanic color palettes.

Trees are spawned to improve the immersion in areas with a high level of detail. They are added as children of the terrain mesh and removed when the LOD falls under a certain level. Two types of trees are used: palm trees for beach biomes and normal trees for other biomes. Tree density varies by altitude, with higher densities in forest biomes and none at high elevations, simulating mountainous regions. To add variety, tree size and rotation are randomized. Tree positions are calculated by selecting a random spot on the quad's plane. Instead of moving trees when LOD changes, their trunks are extended in the model, ensuring they stay grounded and eliminating the need for recalculation.

The sky is enhanced with several objects for added immersion, including a sun, moon, and clouds. The sun is represented as a distant yellow sphere, with a directional light that mimics sunlight at large distances. This method avoids the performance cost of spotlight calculations while maintaining visual accuracy for shadows. The moon orbits the planet like the sun but is smaller, rotates faster, and is positioned closer. It reflects sunlight, allowing for the depiction of lunar phases. Clouds are rendered as billboards with textures by cloud spawners, which are placed at random positions around the sphere. Cloud density and size is randomized to create a diverse cloudscape.

To create an immersive planetary atmosphere, two main effects are implemented: the halo effect and the atmosphere visibility. When viewed from space, a halo effect is rendered around the planet. This is achieved by a shader applied to a mesh slightly larger than the planet itself. The halo's visibility and intensity are modulated by the viewing angle relative to the sun and the planet's surface normal, becoming most prominent around the limb of the planet, particularly on its shadowed side, to simulate scattered light. When the player is within the atmosphere, a different shader dynamically colors the sky. The atmosphere transitions from blue on the sunny side to dark on the other side with a reddish hue in between, simulating the color changes. This effect is achieved through a shader that adjusts color and alpha values based on the position of the player relative to the sun and the planet.

\section{Comparative Study}

To evaluate the perceptual impact of these procedural planet generators, a comparative study was conducted. This study aimed to assess the user experiences across four environments: two featuring planets from the algorithms developed in this work (the ``Simple Planet'' and the ``Minecraft-inspired Planet''). For these, a new, unique planet instance was procedurally generated for each participant. The other two environments were existing, publicly available planet generators. The first is ``Poor Man's Sky'' by jfc3~\cite{PoorMansSky}, which features a first-person camera, a universe with a planet, a star skybox, and basic terrain generation. The second external project is ``ProceduralPlanetGodot'' by Athillion~\cite{AthillionGithub}. It is a procedural planet generator that includes a planet with an interesting ocean shader, a sun, stars, and first-person movement, heavily inspired by the procedural planet generation series of Sebastian Lague on YouTube~\cite{YouTubeTutorialSebastianLague}.

\begin{figure}[hbt!]
  \centering
  \begin{subfigure}[b]{0.48\textwidth}
    \includegraphics[width=\textwidth]{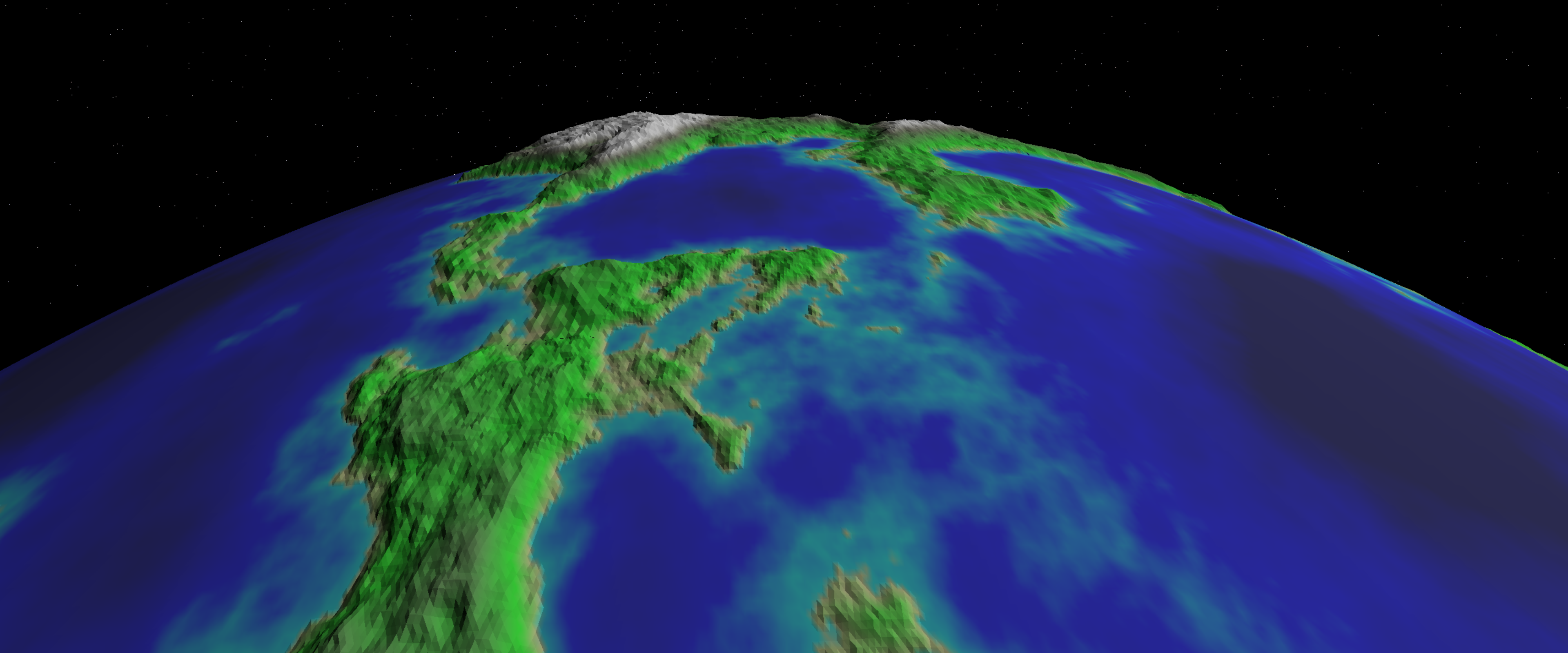}
    \caption{The ``Poor Man's Sky'' planet.}
    \label{fig:poor_mans_sky_visual}
  \end{subfigure}
  \hfill
  \begin{subfigure}[b]{0.48\textwidth}
    \includegraphics[width=\textwidth]{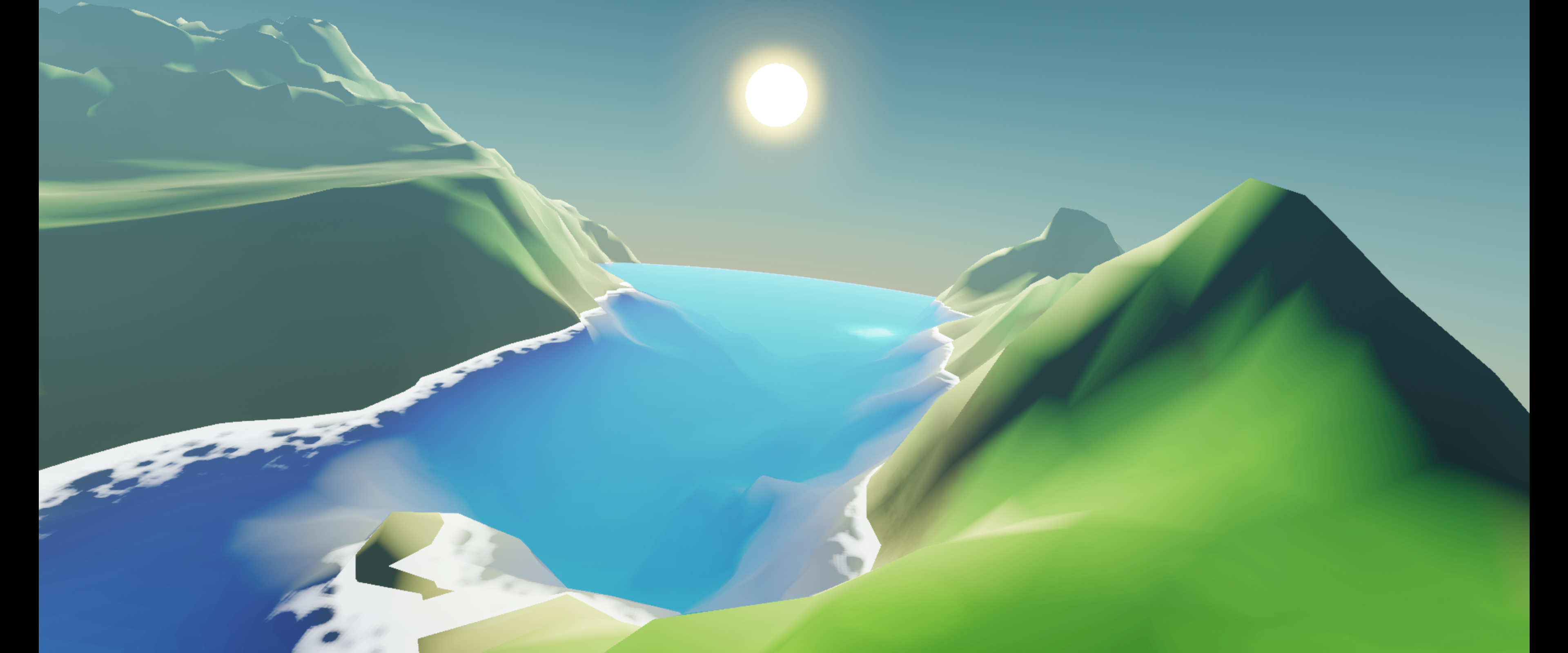} 
    \caption{The ``ProceduralPlanetGodot'' planet.}
    \label{fig:procedural_planet_visual}
  \end{subfigure}
  \caption{Visual examples from the player's perspective of the procedurally generated planets developed in this project. Both images show the in-game rocket model used for exploration.}
  \label{fig:our_generated_planets}
\end{figure}

To compare these four worlds, a questionnaire was developed with a focus on measuring immersion and the experiences of participants during world exploration. The questions chosen are inspired by the ``Igroup Presence Questionnaire''~\cite{Ipq} and the ``Presence Questionnaire'' by Witmer and Singer~\cite{PresenceQuestionnaire} and grouped in the topics \textit{Immersion and Presence}, \textit{Exploration and Fun}, \textit{Terrain Features and Realism}, and \textit{Movement and Controls}. For this exploratory study, a total of 15 participants were recruited with ages ranging from 13 to 58 years with a median age of 26. The group consisted of 11 male and 4 female participants with no individuals identifying as other genders. 

Initially, participants were asked to complete a demographic section of a printed questionnaire, covering age, gender and gaming frequency. The main study involved experiencing four planetary worlds. To mitigate order effects, the sequence in which each participant experienced these four environments was randomized at the beginning of their session using a Python script. 
This script assigned one of the four projects to an identifier (e.g ``Planet 1'') on the questionnaire. For each planetary environment in their randomized sequence, the researcher set up the corresponding application. Participants were given brief instructions on the controls for that specific environment. They were then instructed to freely explore the world for as long as necessary. The primary instruction was to "get a good sense of the world, the terrain and how it feels to move around." 
Immediately after completing the exploration of one planetary environment, participants filled out the corresponding questionnaire section for that planet before proceeding to the next environment in their sequence. This process was repeated until all four environments had been experienced and evaluated. The entire session for each participant lasted between 30 minutes to 1 hour.

Data from the completed paper questionnaires was manually transcribed into a digital spreadsheet, with each row representing a participant's complete set of responses. For the 12-item questionnaire, responses to the two negatively framed questions were reverse-scored to ensure that higher scores consistently indicate more positive perception. For each of the planets, the mean and standard deviation of each question was calculated as well as the mean and standard deviation for each of the predefined question categories.
To explore overall differences between the four environments within each category, one-way between-subjects ANOVAs were conducted. 

\section{Results}

\begin{figure}[hbt]
  \centering
  \includegraphics[width=\columnwidth]{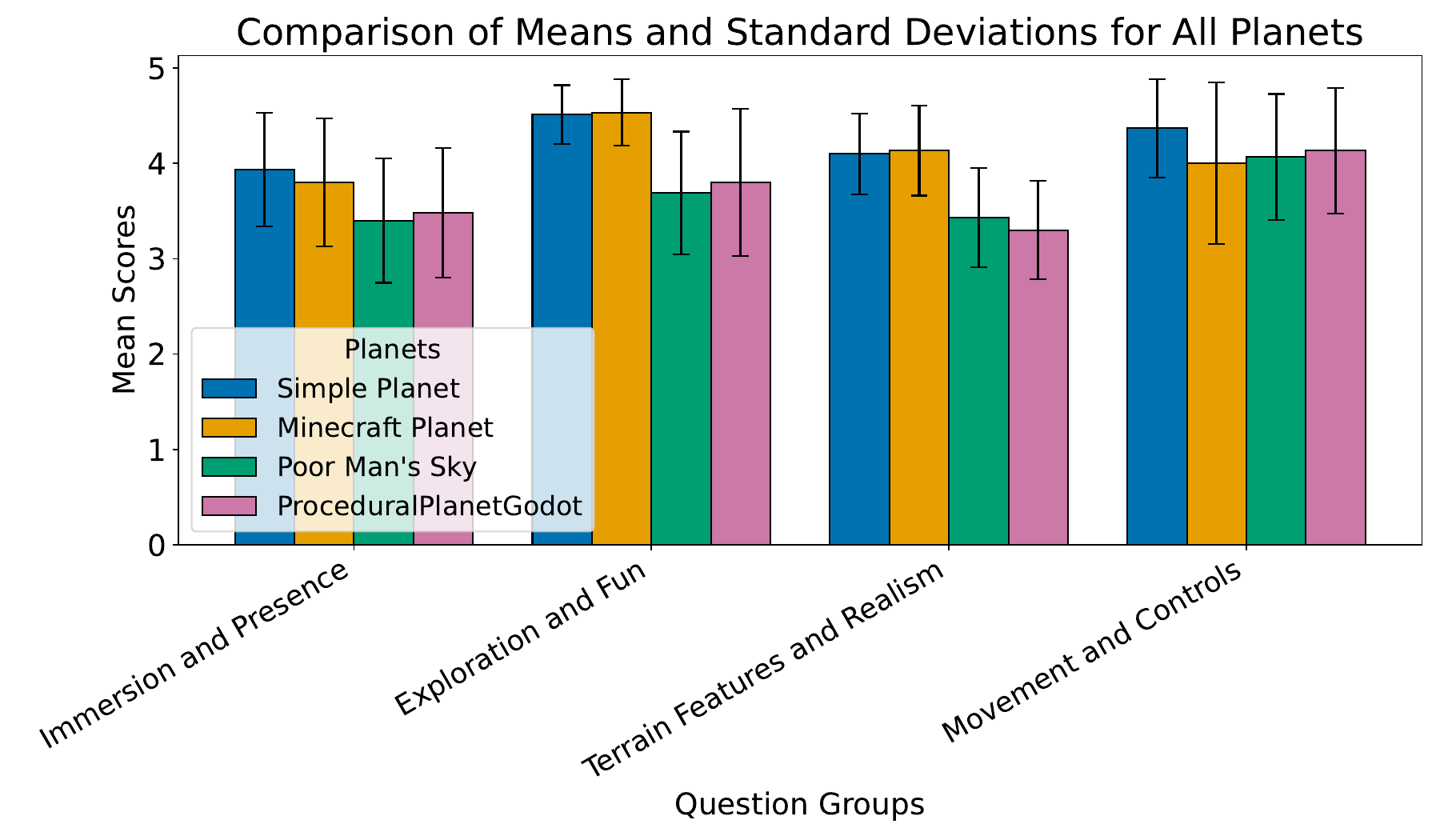}
  \caption{Mean scores (and standard deviations as error bars) for the four perceptual categories across the four planet conditions}
  \label{comparisonMeanStd}
\end{figure}

This section presents participant ratings (N=15) across four perceptual categories. The study compared two planetary generation approaches developed for this work, the ``Simple Planet'' (SP) and the ``Minecraft-inspired Planet'' (MP), against two existing reference projects: ``Poor Man's Sky'' (PMS) and ``ProceduralPlanetGodot'' (PPG). All ratings were made on a 5-point Likert scale (1=Strongly Disagree, 5=Strongly Agree), where higher scores indicate more positive perceptions. The mean scores and standard deviations for each of the four predefined categories across these four planetary environments are visualized in Figure~\ref{comparisonMeanStd}. The one-way ANOVA results used to examine potential statistically significant differences among the planet conditions for each category are summarized in Table~\ref{tab:ANOVAresults}

\begin{table}[hbt]
\centering
\caption{One-way ANOVA Results for Differences Between Planet Conditions}
\begin{tabular}{|l|l|l|l|}
\hline
\textbf{Category}              & \textbf{P-value}  \\ \hline
Immersion and Presence         & 0.0013            \\ \hline
Exploration and Fun            & 0.0043            \\ \hline
Terrain Features and Realism   & 0.0380            \\ \hline
Movement and Controls          & 0.0194            \\ \hline
\end{tabular}
\label{tab:ANOVAresults}
\end{table}

Results are shown in Table~\ref{tab:ANOVAresults}. 
For ``Immersion and Presence,'' the Simple Planet (SP) condition received the highest mean rating (M = 3.93, SD = 0.59), followed by the Minecraft-inspired Planet (MP) (M = 3.8, SD = 0.67). Both ProceduralPlanetGodot (PPG) (M = 3.48, SD = 0.68) and Poor Man's Sky (PMS) (M = 3.4, SD = 0.65) received lower mean scores in this category. The ANOVA indicated a statistically significant overall difference among the conditions for this category 
($p = .0013$).

Turning to ``Exploration and Fun,'' MP (M = 4.53, SD = 0.34) and SP (M = 4.51, SD = 0.31) scored highest, with PMS (M = 3.8, SD = 0.77) and PPG (M = 3.69, SD = 0.64) rated lower. This category also showed a significant overall difference 
($p = .0043$).

Regarding ``Terrain Features and Realism,'' the MP condition again garnered the highest mean rating (M = 4.13, SD = 0.47), with the SP condition also receiving high scores (M = 4.1, SD = 0.42). The simpler terrains of PMS (M = 3.43, SD = 0.52) and PPG (M = 3.3, SD = 0.52) resulted in lower mean scores for this category. The ANOVA indicated a statistically significant overall difference 
($p = .0380$).

For ``Movement and Controls,'' the SP condition was rated highest on average (M = 4.37, SD = 0.52). The MP condition showed a slightly lower mean and more variability (M = 4, SD = 0.85) in its ratings. Both reference projects, PPG (M = 4.13, SD = 0.66) and PMS (M = 4.07, SD = 0.66), received mean scores in this category that were comparable to or slightly higher than the MP condition. The ANOVA also suggested an overall statistically significant difference 
($p = .0194$).

Generally, the planets developed in this project (SP and MP) received higher mean scores across most categories compared to the reference projects (PMS and PPG). For instance, the visually engaging terrains and features like detailed textures, trees, and atmospheric effects in SP and MP likely contributed to these higher ratings. The more complex generation of MP appeared to enhance exploration and perceived realism, though this was sometimes offset by longer loading times, which may have contributed to its slightly lower and more variable scores in ``Movement and Controls'' compared to SP, and even compared to PMS and PPG in that specific category. 

\section{Conclusion}

This paper presented the design, implementation, and comparative user evaluation of two distinct procedural planet generation approaches within the Godot engine. The first, the ``Simple Planet'' (SP), utilized Fractal Brownian Motion with Perlin Noise for terrain, while the second, the ``Minecraft-inspired Planet'' (MP), adapted layered noise techniques for more complex spherical surfaces, both featuring quadtree-based LOD, texturing, and other details to enhance realism. A user study involving 15 participants compared unique instances generated by SP and MP against two existing reference projects across four categories. The findings from this exploratory study indicated that, within this participant group, planets generated using our custom algorithms (SP and MP) generally received more favorable ratings, particularly in aspects of immersion, exploration, and terrain realism. The MP, with its more intricate terrain, often scored highest, though its performance characteristics sometimes impacted ratings in areas like movement and controls compared to the SP. 

\bibliographystyle{splncs04}
\bibliography{bibliography}

\begin{thebibliography}{10}
\providecommand{\url}[1]{\texttt{#1}}
\providecommand{\urlprefix}{URL }
\providecommand{\doi}[1]{https://doi.org/#1}

\bibitem{AthillionGithub}
{Athillion}: Proceduralplanetgodot (2024), \url{https://github.com/athillion/ProceduralPlanetGodot/}, accessed: 2024-09-23

\bibitem{ProceduralContentGenerationTechniques}
Carli, D., Bevilacqua, F., Pozzer, C., d'Ornellas, M.: A survey of procedural content generation techniques suitable to game development. In: A Survey of Procedural Content Generation Techniques Suitable to Game Development. pp. 26--35 (11 2011). \doi{10.1109/SBGAMES.2011.15}, accessed: 2024-09-26

\bibitem{godotCompile}
{Godot Engine}: Compiling for windows - godot engine (2024), \url{https://docs.godotengine.org/en/stable/contributing/development/compiling/compiling_for_windows.html}, accessed: 2024-09-26

\bibitem{Ipq}
{igroup}: igroup presence questionnaire (ipq) overview, \url{https://www.igroup.org/pq/ipq/index.php}, accessed: 2024-09-23

\bibitem{PoorMansSky}
{jfc3}: Poor man's sky, \url{https://jfc3.itch.io/pms/}, accessed: 2024-09-23

\bibitem{ProceduralNoiseFunctions}
Lagae, A., Lefebvre, S., Cook, R., DeRose, T., Drettakis, G., Ebert, D., Lewis, J., Perlin, K., Zwicker, M.: {A Survey of Procedural Noise Functions}. Computer Graphics Forum  (2010). \doi{10.1111/j.1467-8659.2010.01827.x}, accessed: 2024-10-10

\bibitem{YouTubeTutorialSebastianLague}
Lague, S.: Procedural planet generation. YouTube Playlist (2018), \url{https://www.youtube.com/playlist?list=PLFt_AvWsXl0cONs3T0By4puYy6GM22ko8}, accessed: September 29, 2024

\bibitem{FBMTerrain}
Li, Q.Z., Gao, X.R.: Generation and visualization of 3-d realistic natural terrain based on fbm. In: Fourth International Conference on Image and Graphics (ICIG 2007). pp. 915--919 (2007). \doi{10.1109/ICIG.2007.83}, accessed: 2024-09-26

\bibitem{LargeScale3DTerrain}
Liang, Z., Liu, D., Zhou, M.: Research on large scale 3d terrain generation. In: 2014 IEEE 17th International Conference on Computational Science and Engineering. pp. 1827--1832 (2014). \doi{10.1109/CSE.2014.335}, accessed: 2024-09-26

\bibitem{MinecraftWiki}
Minecraft: World generation (2024), \url{https://minecraft.wiki/w/World_generation/}, accessed: September 29, 2024

\bibitem{Minecraft}
Minecraft: Minecraft (2025), \url{https://www.minecraft.net/de-de}, accessed: January 10, 2025

\bibitem{NoMansSky}
{No Man's Sky}: No man's sky, \url{https://www.nomanssky.com/}, accessed: 2024-09-23

\bibitem{Perlin1985}
Perlin, K.: An image synthesizer. SIGGRAPH Comput. Graph.  \textbf{19}(3),  287–296 (Jul 1985). \doi{10.1145/325165.325247}, \url{https://doi.org/10.1145/325165.325247}

\bibitem{ProceduralPlanning}
Roumpani, F.: Procedural cities as active simulators for planning. Urban Planning  \textbf{7}(2),  321--329 (2022). \doi{10.17645/up.v7i2.5209}, \url{https://www.cogitatiopress.com/urbanplanning/article/view/5209}

\bibitem{YouTubeTutorialSimonDev}
SimonDev: 3d world generation (javascript). YouTube Playlist (2021), \url{https://www.youtube.com/playlist?list=PLRL3Z3lpLmH3PNGZuDNf2WXnLTHpN9hXy}, accessed: September 29, 2024

\bibitem{ProceduralGenerationBasics}
Togelius, J., Champandard, A., Lanzi, P.L., Mateas, M., Paiva, A., Preuss, M., Stanley, K.: Procedural content generation: goals, challenges and actionable steps. Dagstuhl Follow-Ups  \textbf{6},  61--75 (01 2013), accessed: 2025-01-11

\bibitem{PresenceQuestionnaire}
{UQO Cyberpsychology Lab, Witmer \& Singer}: Presence questionnaire (2004), \url{https://marketinginvolvement.wordpress.com/wp-content/uploads/2013/12/pq-presence-questionnaire.pdf}, accessed: 2024-09-23, Revised by the UQO Cyberpsychology Lab (2004)

\bibitem{MinecraftTerrainGeneration}
Zucconi, A.: The world generation of minecraft. Alan Zucconi Blog  (2022), \url{https://www.alanzucconi.com/2022/06/05/minecraft-world-generation/}, accessed: September 29, 2024

\end{thebibliography}

\end{document}